\documentclass[5p,times,twocolumn]{elsarticle}

\usepackage{amssymb}

\usepackage[figuresright]{rotating}

\begin{document}

\begin{frontmatter}




\title{Search for time modulations in the decay rate of $^{40}$K and $^{232}$Th}


\author[mi]{E.~Bellotti}
\author[pd]{C.~Broggini\corref{cor1}}
\ead{broggini@pd.infn.it}
\author[lngs]{G.~Di Carlo}
\author[lngs]{M.~Laubenstein}
\author[pd]{R.~Menegazzo}
\author[pd]{M.~Pietroni}

\address[mi]{Universit\`{a} degli Studi di Milano Bicocca and Istituto Nazionale di Fisica Nucleare, Sezione di Milano, Milano, Italy}
\address[pd]{Istituto Nazionale di Fisica Nucleare, Sezione di Padova, Padova, Italy}
\address[lngs]{Istituto Nazionale di Fisica Nucleare, Laboratori Nazionali del Gran Sasso, Assergi (AQ), Italy}

\cortext[cor1]{Corresponding author: Tel. +39 049 9677209}

\begin{abstract}
Time modulations at per mil level have been reported to take place in the decay constant of about 15 nuclei with period of one year (most cases) but also of about one month or one day. In this paper we give the results of the activity measurement of a $^{40}$K source  and a $^{232}$Th one. The two experiments have been done at the Gran Sasso Laboratory during a period of about 500 days, above ground ($^{40}$K) and underground ($^{232}$Th) with a target sensitivity of a few parts over 10$^{5}$. We also give the results of the activity measurement at the time of the X-class solar flares which took place in May 2013. Briefly, our measurements do not show any evidence of unexpected time dependence in the decay rate of $^{40}$K and $^{232}$Th.
\end{abstract}

\end{frontmatter}


\vspace*{-0.5cm}

\section{Introduction}
\label{}
A possible time dependence of the radioactive nuclei decay constant has been searched for since the beginning of the science of radioactivity.
For instance, in the Ph.D. of M. Curie \cite{Cur03} one can already find the negative results of the search for a difference in the radioactivity of uranium between midday and midnight.
Recently, the interest in the subject has been renewed \cite{Jen09} since
various experiments have reported evidence of such an effect with period of one year (most cases) but also of about one month or one day.
In particular, in \cite{Sie98} the constancy of the activity of a $^{226}$Ra source has been measured with an ionization chamber.
An annual modulation of amplitude 0.15$\%$, having the maximum in February and the minimum in August, has been observed. The activity of a $^{152}$Eu source has also been measured by means of a Ge(Li) detector and an even larger
annual modulation (0.5$\%$) has been detected.
Alburger and Collaborators  \cite{Alb86}  measured the half-life of $^{32}$Si, which is interesting for many applications in Earth Science. Data, collected in four years, show an evident annual modulation,
with an amplitude of about 0.1$\%$ .
In \cite{Jen09,Fis11} the existence of new and unknown particle interaction has been advocated to explain the yearly variation in the activity of radioactive sources \cite{Sie98,Alb86}.
The Authors correlate these variations to the Sun-Earth distance. On the other hand, the possibility for anti-neutrinos affecting the $\beta^{+}$ decay has been excluded in a recent reactor experiment \cite{Mei11}.

The paper by Jenkins and Collaborators \cite{Jen09} triggered a renewed interest in the subject. Old and recent data have been analyzed, or reanalyzed, to search for periodic and sporadic variations with time.
Some of the measurements and analyses confirm the existence of oscillations \cite{Par10,Jav10,Vep12} whereas others contradict this hypothesis \cite{Har11,Coo08,Nor08,Sch10,Bik13}.
For instance, \cite{Par10} presents the time dependence of  the counting rate  for $^{60}$Co, $^{90}$Sr and $^{90}$Y sources, measured with Geiger M\"{u}ller detectors and for a $^{239}$Pu source,
measured with silicon detectors.
While beta sources show an annual (and monthly) variation with amplitude of about 0.3$\%$, the count rate from the Pu source is fairly constant. The most updated list of experimental results on different nuclei showing time-dependent decay rate is given in \cite{Jen12}.

Since we believed that more dedicated experiments were needed to carefully investigate the time dependence of the decay constant, two years ago
we performed an experiment underground at the Gran Sasso Laboratory (LNGS) with a $^{137}$Cs source and a heavily shielded Germanium detector \cite{Bel12}.
Time dependence in the order of 0.2$\%$ were previously reported  for $^{137}$Cs \cite{Bau01}.
Results of our experiment have already been published. Very briefly,
any oscillation with period between 6 hours and 1 year and with amplitude larger
than 9.6$\cdot$10$^{-5}$ has been excluded at 95$\%$ confidence level (C.L.). In particular, for an oscillation period of 1 year a limit of 8.5$\cdot$10$^{-5}$ at 95$\%$ C.L.
on the maximum allowed amplitude has been set independently of the phase.

In the next section we describe our methodological approach, before giving in section 3 and 4 the results of the two new experiments we performed to study the electron-capture partial decay constant of  $^{40}$K and the decay chain of $^{232}$Th (alpha and beta decay), the former above ground and the latter underground  at LNGS. Before giving conclusions, we describe in section 5 the decay rate of the two radioactive sources in correspondence with the strong X-class solar flares which took place in May 2013.
Finally, we show in the Appendix that the size of a possible modulation induced by the coupling with a scalar field sourced by the Sun (one of the few mechanisms introduced to explain the effect) could be only at the level of 10$^{-12}$, at most.

\section{The methodological approach}
\label{}
The general aim of our dedicated experiments is to reach a sensitivity better than 1 part over 10000 (1$\sigma$ level) on a
possible annual modulation.
For this a rate of the order of $10^8$ events/day
is needed. Note that this is true as far as the statistical uncertainty only is concerned,
without including possible systematical effects
that can only worsen the sensitivity.

Gamma spectroscopy offers a powerful and simple method to investigate
the stability of nuclear decay rates. As a matter of fact, there is a
wide choice of radioactive isotopes, half lives, gamma ray energies, source dimensions
and physical forms (point-like or extended). Moreover, with a
gamma ray detector of good energy resolution, like germanium or sodium iodide crystals, it is
possible to identify the emitters and in most cases also the origin of the
background.

The typical experimental configuration that we adopted is the
following: a radioactive source that illuminates a detector,
an electronic chain that processes the signal, a multichannel analyzer
that produces an energy spectrum, saved by an acquisition system at
fixed intervals of time. The source/detector assembly is surrounded by
a heavy lead shielding in order to suppress the external gamma background.
Cosmic rays are an unavoidable background source that cannot be
suppressed with any realistic amount of lead or other shielding material. In addition,
their flux is not constant in time, giving a serious limit
to the investigation of subtle variations in nuclear decay rates as the ones
we are interested in.
The solution we choose to avoid such a background in the $^{232}$Th experiment is to install the
experiment deep underground in the Gran Sasso Laboratory. The mountain
shield suppresses the muon and neutron flux by six and three orders of magnitude, respectively,
as compared to the above ground.

In the measurements described in the following sections we consider
the entire energy spectrum and not only the full energy peaks. This because
we want to avoid any inaccuracy coming from the fitting procedure and we also want
to increase the total rate in order to improve the statistics. This procedure
requires the definition of lower and upper boundaries. Since
the content of the energy spectrum above $\sim$3 MeV is negligible, the only delicate
point is the stability of the lower threshold that should be low enough to
collect the entire spectrum and high enough to be well above the electronic noise.
Note that if the low energy threshold is sufficiently low and placed inside a flat region of the spectrum then it is also less
sensitive to the variation of the global gain of the electronic chain.
Spectra are collected at fixed intervals (1 hour or 1 solar day), with
the timing for data acquisition provided by the internal quartz
oscillator of the acquisition card. Its precision and stability
(better than 10 ppm/year) are enough for our purposes.

Sources can have different configurations, point-like
or extended.
In the case of point-like sources it is essential to keep
its position  fixed as well as the distance between the source itself and the detector.
As a matter of fact, a variation of only 1 micron in the distance gives rise in one of our
typical set-ups to a change in the detection efficiency of about 5$\cdot$10$^{-5}$ .
In addition to the mechanical stability of the source support, it is also necessary to
avoid large temperature and atmospheric pressure fluctuations, that can
modify the source-detector distance. The typical length variation
with temperature in metals is of the order of
20 ppm per degree. As a consequence,
a moderate thermal stabilization is needed to avoid
spurious effects on the rate. The location of our experimental
setups in underground laboratory or in the basement of above ground
buildings represents a well acceptable solution. In any case, temperature
and pressure are monitored. On the other hand, the selection of an extended source, if possible, makes all the above discussed requests
much less severe.

Once collected, spectrum data are averaged over a period of 1 day (or
larger if necessary) and then fitted with the expected curve,
normally an exponential, unless the source has not yet reached the secular equilibrium. The presence of
effects due to unknown systematics and/or non standard behaviors
can be indicated by bad chi squared values of the fit.
Finally, time modulations of the rate are searched
for with the Fourier transform method applied to the residuals or with the minimization of the chi squared  fit of the residuals with a cosine function of time. As a matter of fact, Fourier transform can be rigorously applied only when searching for periods
significantly smaller than the counting time. In particular, we have chosen a limit equal to one third of the run time.

\section{The $^{40}$K experiment}
\label{}
The detector, a 3"x3" NaI crystal, is installed above ground, in the basement of one of the LNGS buildings.
The source is made by about 16 kg of potassium bicarbonate powder (KHCO$_{3}$, corresponding to 6.24 kg of natural potassium) contained inside a stainless steel box placed around the detector. The whole set-up is shielded by at least 10 cm of lead and the electronic signals are processed by an Ortec Digibase (shaping time: 0.75 $\mu$s).

\begin{figure}[h]
\centerline{
\includegraphics[width=0.5\textwidth,angle=0]{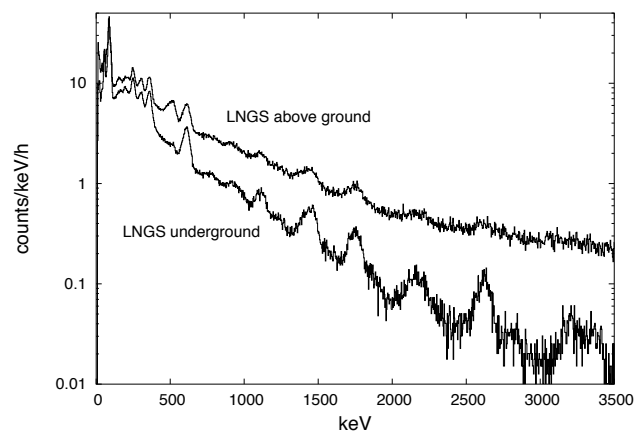}
}
\vskip -5mm
\caption{Spectra taken with a lead shielded 3"x3" NaI crystal placed first close to the $^{40}$K set-up and then underground in the Gran Sasso laboratory, averaged over 5 days. The 1024 ADC channels cover the energy range between 15 keV and 3500 keV.
}
\label{ampute}
\end{figure}

The intrinsic background, measured in similar conditions, but without the powder inside the box, gives a rate of 2.4 s$^{-1}$ in the energy window 15-3500 keV.
The contribution to the background due to cosmic rays has been measured by comparing the spectra obtained with a lead shielded 3"x3" NaI crystal placed first close to the $^{40}$K set-up and then underground in the Gran Sasso laboratory (Figure 1).
It amounts to 0.997(5) s$^{-1}$,
to be compared with 788 s$^{-1}$ due to the potassium source.The underground spectrum is below the above ground one at any energy, in particular the line at 511 keV due to  positron annihilation is absent underground.
We observe that electrons from beta decay give rise to
bremsstrahlung photons which can be detected by the sodium iodide crystal in addition to
the ones from electron capture decay. We have evaluated by Monte Carlo simulation that the bremsstrahlung photons,
because of the absorption in the 1.5 mm thick stainless steel box containing the salt and in the 0.5 mm
thick aluminum body around the crystal, give a contribution of, at most, 5$\%$ to the rate.
This contribution is confined to the gamma energy region below 300 keV.

Spectra are stored once per hour with a relative dead time of 3.54$\cdot$10 $^{-3}$, which has a fluctuation of
 1.5$\cdot$10 $^{-5}$, at most.
During the 700 days of data taking we measured a peak position shift (proportional to the channel number) of at most
14 channels for the $^{40}$K peak at 1461 keV energy. This shift is due both to the high voltage drift and to the temperature variation (at most 6 degrees, slowly changing with time).

\begin{figure}[h]
\centerline{
\includegraphics[width=0.5\textwidth,angle=0]{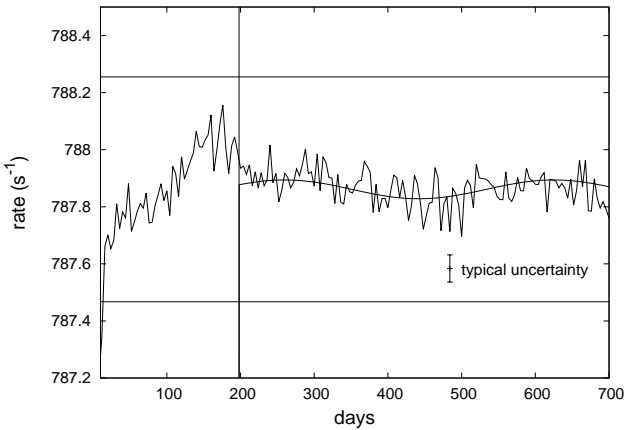}
}
\vskip -6mm
\caption{Measured rate of the $^{40}$K source averaged over 1 day. Day zero is November 30th, 2011. The vertical line marks the
end of the stabilization time for the potassium bicarbonate powder. The $\chi^2$ minimization gives the cosine curve. The two horizontal
lines are drawn at $\pm$5$\cdot$10$^{-4}$ from the average.
}
\label{ampute}
\end{figure}

The rate is shown in Figure 2.
We see that there is an increase up to 0.3 s$^{-1}$, corresponding to less than 4 $\cdot$10$^{-4}$,
during the first 4 months of the experiment, followed by a decrease of 0.2 s$^{-1}$ in the next
2 months. Afterwards the rate stays constant in time for more than 500 days.
The behavior in the first 200 days of data taking is due, in our opinion, to two processes in
competition: the settlement of the powder inside the steel box, that increase the detection
efficiency, and the capture of water by the potassium salt, slightly hygroscopic, that
decreases the efficiency due to the increase of the average density of the extended source and of its
self absorption (the box containing the powder is closed but not hermetically sealed). As a matter of fact,
when we opened the box containing the salt at the end of the experiment we saw that the
salt, originally a crystalline powder, had become a unique solid body with a volume reduction of about
1.5$\%$. This volume reduction, together with the water absorption, gave rise to the detection efficiency
change responsible for the non constant rate during the first 200 days of the experiment.
In the following analysis we refer only to the data taken starting from day 200.
We do not consider to do any dead time correction to the data but we do apply a correction to the
content of the first bin due to the peak position shift (0.005 $\%$ at most).
The rate as function of time, when fitted with a constant and using as uncertainty the linear sum of the
Poisson fluctuation (1.2$\cdot$10$^{-4}$ relative uncertainty) plus the fluctuation of the
live time  (7$\cdot$10$^{-6}$ relative uncertainty) gives a chi squared per degree of freedom,
$\chi^2/dof$, of 1.08.

Figure 3 shows the results of the Fourier transform of the residuals, for periods from 3 to 150 days, and of the chi-squared
analysis for longer periods: there is a clear modulation for a period of 1 year with an amplitude of 4.5$\pm$0.8$\cdot$10$^{-5}$
and maximum at August 11th ($\pm$ 13 days). It corresponds to a $\pm$ 3.5$\%$ modulation of the cosmic ray background we measured  and it is well compatible both in size and in phase with the known annual modulation of the cosmic
ray flux at the Earth surface, which we must see because of the sensitivity of our experiment.
The inclusion of this oscillating component into the fit lowers the $\chi^2/dof$ to 1.03.
On the other hand, any oscillation with period different from 1 year and with amplitude larger than 2.6$\cdot$10$^{-5}$ is
excluded at 99.7$\%$ C.L. independently of the phase.

\begin{figure}[h]
\centerline{
\includegraphics[width=0.5\textwidth,angle=0]{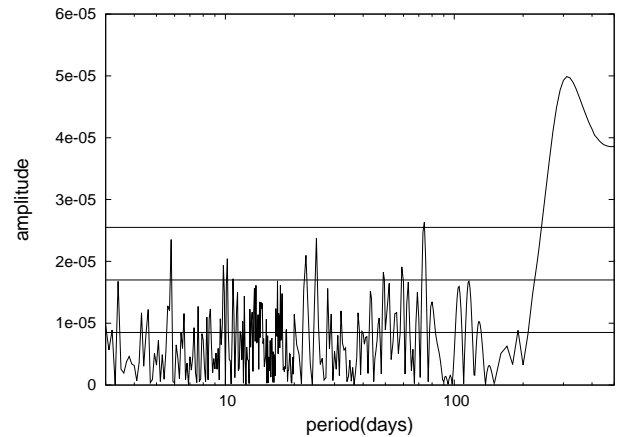}
}
\vskip -6mm
\caption{Potassium source: amplitude of the different periods from the Fourier transform of the residuals and from the chi-squared
analysis. Horizontal lines correspond to 1,2 and 3 standard deviations, obtained with a 10000 time re-shuffling of the residuals or from the
chi-square.
}
\label{ampute}
\end{figure}

If, for sake of completeness, we consider the entire data taking period, including also the first 200 days, we can in any case exclude a maximum
variation of the rate larger than $\pm$ 3 $\cdot$ 10$^{-4}$ without any evident periodicity,
as shown in Figure 2.

\section{The $^{232}$Th experiment}
\label{}
The activity of a Thorium source (alpha and beta decay) is measured with a 3"x3" NaI crystal
installed deep underground in the  STELLA facility (SubTerranean Low Level Assay) of LNGS.
As a consequence, the background due to cosmic rays and neutrons is strongly suppressed.
The source is obtained from an optical lens, made by special glass heavily doped with Thorium Oxide. Note that this technique, used for improving the optical properties of glass, was quite common until
the seventies. The lens is placed close to the crystal housing. Both the lens and the NaI
detector are shielded with at least 15 cm of lead. The electronic chain consists of a PMT base with a preamplifier, a SILENA amplifier (shaping time: 0.5 $\mu$s) and a MCA Canberra Multiport II.

\begin{figure}[h]
\centerline{
\includegraphics[width=0.5\textwidth,angle=0]{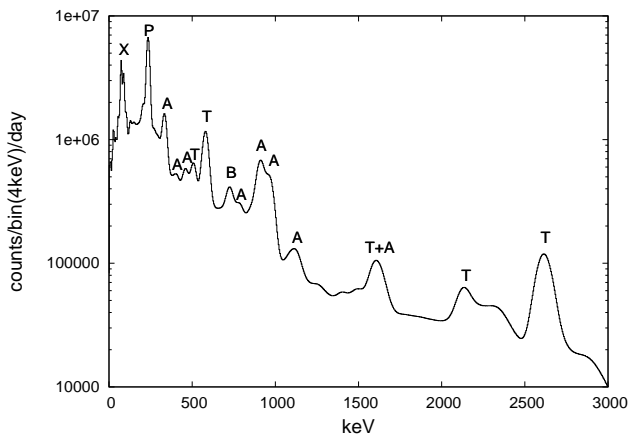}
}
\vskip -6mm
\caption{Spectrum of the Thorium source, averaged over 110 days. Gamma peaks are due to lead X-rays (X) and to the decay
 of $^{212}$Pb (P), $^{228}$Ac (A), $^{212}$Bi (B) and $^{208}$Tl (T). The 1024 ADC channels cover the energy range between 12 keV and 4120 keV.
}
\label{ampute}
\end{figure}

The intrinsic background rate above the 10 keV threshold is of 2.3 s$^{-1}$
(due to $^{40}$K, thorium and uranium chains and lead X-rays) to be compared to 3171 s$^{-1}$ from the source. Spectra (Figure 4) are stored once per day with a dead time of about 1.3 $\%$. During the 480 days of data taking we measured a peak position shift (proportional to the channel number) which is at most
of 15 channels for the $^{208}$Tl peak at 2615 keV energy. The rate is shown in Figure 5.
We see that the chain is not in equilibrium since there is a rate increase of $1.8\cdot10^{-3}$ over a time period of 1 year. The time dependence of the rate is well described by the equations ruling the recovery of secular equilibrium \cite{Arp96}, as shown by the continuous line in Figure 5. This is consistent with the glass of the lens being produced 38 years ago, when the removal of  $^{228}$Ra (isotope of the thorium chain with half-life of 5.75 years) took place.

\begin{figure}[h]
\centerline{
\includegraphics[width=0.5\textwidth,angle=0]{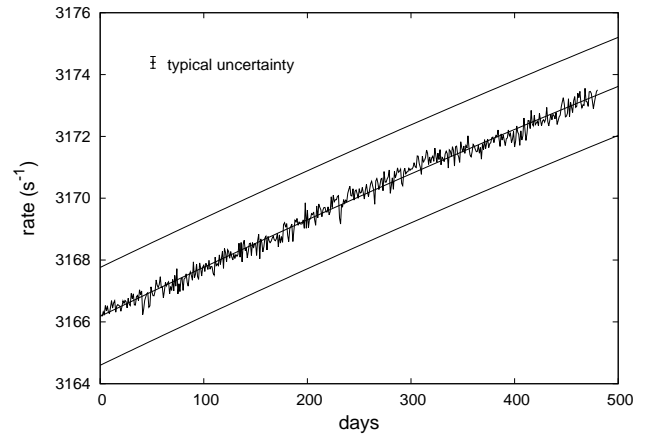}
}
\vskip -6mm
\caption{Rate of the Thorium source averaged over 1 day. Day zero is July 25th, 2012. The continuous line shows the recovery of the secular equilibrium. The two
lines are drawn at $\pm$5$\cdot$10$^{-4}$ from the continuous one.
}
\label{ampute}
\end{figure}

We searched for time modulations without applying any dead time correction to the data. As a matter of fact,  the expected rise of dead time in 1 year (about 2.4$\cdot$10$^{-5}$) due to the non-equilibrium of the chain is not affecting our time modulation search. Figure 6 shows the results of the Fourier transform of the residuals, up to a period of 150 days, and of the chi-squared analysis for longer periods. There is a significant period, 300 days, with amplitude of
4$\cdot$10$^{-5}$. However, the same period is present, with an amplitude of 2$\cdot$10$^{-2}$, in the daily averaged dead time per event, obtained from the multichannel analyzer, which, on the contrary, should stay constant. The modulation of the rate and of the daily averaged dead time per event
have the same period and opposite phase.
We attribute this effect, which is the intrinsic limit of the measurement, to the instabilities of  the RC circuit providing the shaping time of the amplifier. As a matter of fact, a not well fixed shaping time gives rise to a variable duration of the veto signal and, as a consequence, to a variable number of rejected pile-up events. For comparison, the fluctuations of the dead time per event measured in the potassium experiment, where a different electronic chain is used, are smaller by a factor 10 and do not show any periodicity.

\begin{figure}[h]
\centerline{
\includegraphics[width=0.5\textwidth,angle=0]{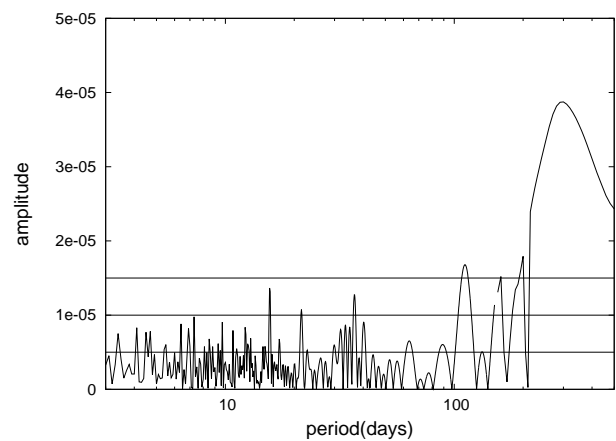}
}
\vskip -6mm
\caption{Thorium source: amplitude of the different periods from the Fourier transform of the residuals and from the chi-squared
analysis. Horizontal lines correspond to 1,2 and 3 standard deviations, obtained with a 10000 time re-shuffling of the residuals or from the
chi-square.
}
\label{ampute}
\end{figure}

Therefore, we consider the 300 day modulation to be a systematic effect and, in spite of the excellent statistical uncertainty of
5$\cdot$10$^{-6}$, we can only exclude any oscillation with period smaller than 1 year and amplitude larger than 4$\cdot$10$^{-5}$, being this the size of our largest systematic uncertainty.

\section{Source activity during the X-class solar flares of May 2013}
\label{}
Solar flares are explosions on the Sun that happen when energy stored in twisted magnetic fields (usually above sunspots) is suddenly released.
This energy, up to one hundredth of the solar luminosity, is released within a few minutes to tens
of minutes. In this interval the plasma is heated to  tens of millions of degrees with a
strong X-ray emission together with electron and proton acceleration (up to several tens and hundreds of MeV, respectively).
Solar flares are classified according to the power of the X-ray flux peak near the Earth as
measured by the GOES-15 geostationary satellite:
X identifies the class of the most powerful ones, with a power at the peak larger than 10$^{-4}$  W/m$^{2}$
(within the X-class there is then a linear scale).

In \cite{Jen06} a significant dip (up to 4$\cdot$10$^{-3}$, $\sim$7 $\sigma$ effect) in the count rate,
averaged on a time interval of 4 hours, has been observed in the activity of a
$\sim$1 $\mu$Ci source of $^{54}$Mn (electron-capture) in coincidence with the X-class solar flares from December 2nd 2006 to January 1st 2007.
On the other hand, a different experiment with a $\sim$10$^{-3}$ sensitivity,
carried out by Parkhomov \cite{Par10}, did not observe any deviation in the
activity of $^{60}$Co, $^{90}$Sr-Y and $^{239}$Pu sources in coincidence with the same flares.

We have already published results \cite{Bel13} from the data we took with the $^{40}$K, $^{137}$Cs and $^{232}$Th sources during the two strongest solar flares of the years 2011 and 2012
(X5.4 and X6.9). Very briefly, no significant deviations from standard expectation have been observed, with a sensitivity of $3\cdot$10$^{-4}$ per day. Here we give the results obtained by studying the activity of the $^{40}$K and $^{232}$Th sources in correspondence with the 4 X-class solar flares solar which took place between the 13th and the 15th of May 2013 (X1.7, X2.8, X3.2 and X1.2).

Figure 7 shows the data collected in a 20 day window centered on the 14th of May 2013
(the day is given in terms of the Modified Julian Date).
The X-ray peak flux is plotted in linear scale and given in W/m$^2$, in the 0.1-0.8 nm
band measured by the GOES-15 satellite \cite{noaa}. Inside the two bands
are plotted the residuals
of the normalized count rate of the  $^{40}$K and $^{232}$Th sources (i.e. the difference between the measured and expected count rate divided by the measured one), averaged over a period of 1 day.
The error bars are purely statistical. Systematic uncertainties are negligible as compared to the statistical ones during a data taking period of
a few days only.
For the $^{232}$Th data a linear trend (5.7 ppm/day), due to the recovering of the secular equilibrium,
is subtracted.

\begin{figure}[h]
\centerline{
\includegraphics[width=0.5\textwidth,angle=0]{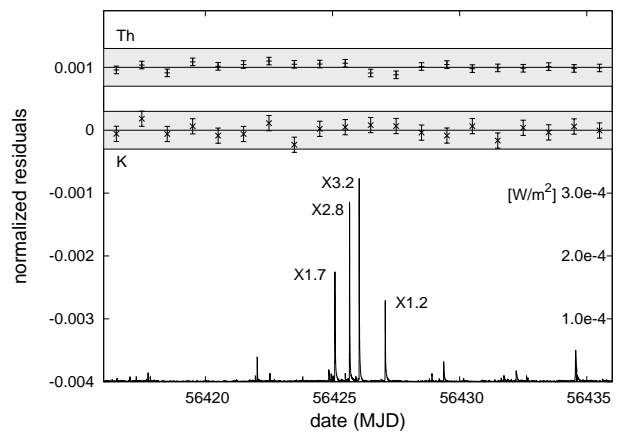}
}
\vskip -6mm
\caption{Residuals, averaged over 1 day, of $^{40}$K and $^{232}$Th data collected around the 14th of May 2013 (56426) and the X-ray flux
measured by the GOES-15 Satellite.
The $^{232}$Th data are vertically displaced by 0.001 for sake of clarity; the right-hand vertical scale
gives the X-ray flux measured in W/m$^2$. The two shaded bands are drawn at $\pm 3\cdot 10^{-4}$
from the expected value.
}
\label{ampute}
\end{figure}

Our data clearly exclude an effect as large as the one reported in \cite{Jen06}, of the order of a few per mil per day and lasting
several days, and confirm the same results as in \cite{Bel13}. In particular,
the maximum effect compatible with our data is smaller than $3\cdot 10^{-4}$ per day
at $95\%$ C.L. for the 4 X-class flares of May 2013. Such a limit is obtained by taking twice the statistical uncertainty.

\section{Conclusions}
\label{}
Recently, various experiments have reported evidence of a time dependence of the radioactive nuclei decay constant with period, in most cases, of one year and amplitude at the per mil level. In particular, the annual modulation has been correlated to the change of the Sun-Earth distance between aphelion and perihelion.
In the Gran Sasso Laboratory we performed two experiments to search for modulations in the electron-capture decay of
$^{40}$K and in the decay chain of $^{232}$Th (alpha and beta decay).

In the decay rate of $^{40}$K we measured a relative annual modulation of  4.5$\pm$0.8$\cdot$10$^{-5}$. However, the $^{40}$K
experiment was located above ground and the measured effect is well compatible with the annual modulation of the cosmic ray flux (3.5$\%$ amplitude of the cosmic ray background and maximum in July-August). Any oscillation with period lower than 1 year and with amplitude larger than 2.6$\cdot$10$^{-5}$ is
excluded at 99.7$\%$  C.L.  independently of the phase.

On the contrary, the $^{232}$Th experiment was located underground. The analysis of the thorium data excludes with a 4$\cdot$10$^{-5}$ sensitivity any oscillation with period  shorter than 1 year independently of the phase.

Finally,  the gamma activity of $^{40}$K and $^{232}$Th
have been measured during the occurrence of the 4 strong X-class solar
flares of May 2013. No significant deviations from expectations have been observed.

\section*{Acknowledgments}
We thank the Director and the staff of LNGS together with Prof. R.Battiston, Dr. C.Cattadori and Dr. L.Pandola for their continuous support.



\appendix
\section{Constraints from variations of fundamental constants}
\label{}

An annual time modulation of nuclear decay constants could be explained by the presence of a new interaction mediated by a scalar field sourced by the Sun \cite{Jen09,Fis11}. In this appendix we discuss this scenario, showing that existing bounds on the Weak Equivalence Principle (WEP) and on the stability of atomic clocks  (for a review, see \cite{Uza02} ) already exclude this explanation.

The orbital motion of the Earth around the Sun would induce an annual modulation of the fundamental parameters entering the nuclear rate $\Gamma$, which would in turn exhibit a variation given by

\begin{equation}
\frac{\Delta \Gamma}{\Gamma}= \sum_i \frac{\partial \log \Gamma}{\partial \log C_i} \Delta \log C_i\,,
\label{sensi}
\end{equation}
where the $C_i's$ are the fundamental parameters entering the rate. In the following, we will consider the dependence on $\alpha_{em}$, and on the Fermi constant, $G_F$, {\it i.e.} $C_i=\{\alpha_{em},\,G_F\}$.

\subsection{Dependence on $\alpha_{em}$}
Consider the $\alpha$ decay

\begin{equation}
^{A+4}_{Z+2}X\to ^{A}_{Z}Y +^{4}_{2}{\rm He}\,.
\end{equation}

The decay rate is governed of the penetration of the Coulomb barrier described by Gamow theory, and well approximated by

\begin{equation}
\Gamma \simeq \Lambda(\alpha_{em},v) e^{-4\pi Z\alpha_{em}c/v},
\end{equation}

where $v$ is the escape velocity of the $\alpha$ particle and $\Lambda$ is a function that depends mildly on $\alpha_{em}$ and $v$. It follows that the variation of the decay rate with respect to the fine structure constant is well approximated by \cite{Uza02}

$$
s\equiv \frac{d \log\Gamma}{d\log \alpha_{em}} \simeq 4 \pi Z \frac{c}{v} \alpha_{em}
\left\{\left(\frac{0.3\,{ \mathrm{MeV}}}{\Delta E} \right)f(A,Z) -1 \right\}\,,
$$

where $$f(A,Z) = (Z+2)(Z+1)(A+4)^{-1/3}-Z(Z-1)A^{-1/3}$$ and $\Delta E$ is the decay energy. As an example, for the $^{238}_{92}$U $\alpha$ decay ($\Delta E= 4.27 \,\mathrm{MeV}$) the sensitivity is $s\simeq 540$.

In a $\beta$ decay the sensitivity to $\alpha_{em}$ can be written as \cite{Uza02}

\begin{equation}
s\equiv \frac{d \log\Gamma}{d\log \alpha_{em}} \simeq p\frac{d\log \Delta E}{d \log \alpha_{em}}\,,
\end{equation}

where $p=2 +\sqrt{1-\alpha_{em}^2 Z^2}$ and

\begin{equation}
\frac{d\log \Delta E}{d \log \alpha_{em}} =\left(\frac{0.6\,\mathrm{MeV}}{\Delta E}\right)\,.
\end{equation}

For the $\beta$ decay of $^{87}_{37}$Rb, $(\Delta E=0.275\,\mathrm{MeV})$, one gets $s\simeq -180$.
\label{labo}
The strongest constraints on the stability of $\alpha_{em}$ from laboratory experiments comes from the comparison of Rubidium and Cesium atomic clocks. Since they are working on different transitions, the parametric dependences on $\alpha_{em}$ of the frequency of the clocks are different, and the absence of drift poses upper bounds on the variation of $\alpha_{em}$. The clocks were observed over a period of 20 months, leading to the bound \cite{Pre95}

\begin{equation}
\frac{\dot{\alpha}_{em}}{\alpha_{em}}= (4.2 \pm 6.9) \cdot 10^{-15}\;{\mathrm{yr}}^{-1}\,.
\end{equation}

Using eq.~\ref{sensi} and the $O(10^2)$  sensitivities of the previous subsection, we get an extremely suppressed value

\begin{equation}
\frac{\Delta \Gamma}{\Gamma} < O(10^{-13})\,.
\end{equation}

\subsection{Dependence on $G_F$}
The rate for $\beta$ decay is proportional to $G_F^2$, therefore the sensitivity in this case is simply given by
\begin{equation}
s_W \equiv \frac{d \log\Gamma}{d\log G_F} =2\,.
\label{sw}
\end{equation}
A variation of $G_F$ can be induced by varying the Higgs expectation value (VEV),  $v_{W} = \left(\sqrt{2}\, G_F\right)^{-1/2}$. A variation of the Higgs VEV, would manifest itself in an annual variation of the electron to proton mass ratio $\mu$, and consequently of spectral lines, of the same amount. Variations of this kind are bounded by atomic clocks, as discussed in the previous subsection. The best bound  comes from comparing cesium atomic clocks with superconducting microwave cavities oscillator. The frequency of the cavity-controlled oscillators was compared during 10 days to that of cesium. The relative drift rate was $< 10^{-14}\;{\mathrm{day}}^{-1} $ \cite{Uza02}, which, extrapolated to an annual timescale, gives $\Delta G_F/G_F <O (10^{-12})$ and then, using \ref{sw},

\begin{equation}
\frac{\Delta \Gamma}{\Gamma} < O(10^{-12}).
\end{equation}

Our analysis clearly excludes a scalar field from the Sun as a possible origin of a modulation of the nuclear decay rate with a per mil amplitude and, as a consequence, a different mechanism would have to be found to explain the reported evidence, if it were confirmed.





\end{document}